\documentclass[conference]{IEEEtran}  %This is for PES
\IEEEoverridecommandlockouts
\usepackage{ifpdf}
\usepackage{cite}
\usepackage{graphicx}
\usepackage{epstopdf}
\usepackage{amssymb}
\usepackage{bm}
\usepackage{amsthm}
\usepackage[ruled,vlined]{algorithm2e}

\ifCLASSOPTIONcompsoc
    \usepackage[caption=false, font=normalsize, labelfont=sf, textfont=sf]{subfig}
\else
\usepackage[caption=false, font=footnotesize]{subfig}
\fi

\ifCLASSINFOpdf
\else
\fi

\usepackage[cmex10]{amsmath}

\usepackage{color}

\begin{document}
\title{Energy Contract Settlements through Automated Negotiation in Residential Cooperatives}
% \title{Negotiating Energy Contracts: An Alternative Form of Marketplace for Residential Energy Cooperatives}
% \title{An Exchange Mechanism for Energy Cooperatives with Flexibility Coordination}

% \author{
% \IEEEauthorblockN{Shantanu Chakraborty, Tim Baarslag and Michael Kaisers}
% \IEEEauthorblockA{Intelligent and Autonomous Systems Group\\
% Centrum Wiskunde \& Informatica, Amsterdam, The Netherlands\\
% \{Shantanu.Chakraborty, T.Baarslag, Michael.Kaisers\}@cwi.nl}
% }

% \author{
% \IEEEauthorblockN{Shantanu Chakraborty, Tim Baarslag and Michael Kaisers}
% \IEEEauthorblockA{Intelligent and Autonomous Systems Group\\
% Centrum Wiskunde \& Informatica, Amsterdam, The Netherlands\\
% \{Shantanu.Chakraborty, T.Baarslag, Michael.Kaisers\}@cwi.nl}
% }

%% To specify the authors when (number of affiliations <= 2)
% \author{
% % \IEEEauthorblockN{Author 1, Author 2, Author 3}
% \IEEEauthorblockA{Intelligent and Autonomous Systems Group\\
% Centrum Wiskunde \& Informatica, Amsterdam, The Netherlands\\
% Shantanu.Chakraborty@cwi.nl}
% }

%% To specify the authors when (number of affiliations > 2)
\author{\IEEEauthorblockN{Shantanu Chakraborty}
\IEEEauthorblockA{The University of Melbourne, Australia.\\
shantanu.chakraborty@unimelb.edu.au}
\and
\IEEEauthorblockN{Tim Baarslag and Michael Kaisers}
\IEEEauthorblockA{Centrum Wiskunde \& Informatica, Amsterdam, The Netherlands.\\
\{T.Baarslag, Michael.Kaisers\}@cwi.nl}
\thanks{This work has been submitted to the IEEE for possible publication. Copyright may be transferred without notice, after which this version may no longer be accessible.}
}

% \author{\IEEEauthorblockN{Shantanu Chakraborty\IEEEauthorrefmark{1},
% Tim Baarslag\IEEEauthorrefmark{2} and Michael Kaisers\IEEEauthorrefmark{2}}
% \IEEEauthorblockA{\IEEEauthorrefmark{1} Department of Engineering and Commissioning \\
% Siemens Netherlands, The Hague, The Netherlands.\\
% Shantanu.Chakraborty@siemens.com}
% \IEEEauthorblockA{\IEEEauthorrefmark{2} Intelligent and Autonomous Systems Group\\
% Centrum Wiskunde \& Informatica, Amsterdam, The Netherlands.\\
% \{T.Baarslag, Michael.Kaisers\}@cwi.nl}}
% \IEEEauthorblockA{\IEEEauthorrefmark{3} Department Name of Organization C\\
% Name of the organization C,
% Address C\\ Emails if wanted}
% \IEEEauthorblockA{\IEEEauthorrefmark{4}Department Name of Organization D\\
% Name of the organization D,
% Address D\\ Emails if wanted}
% }

% make the title area
\maketitle

\begin{abstract}
This paper presents an automated peer-to-peer (P2P) negotiation strategy for settling \emph{energy contracts} among prosumers in a Residential Energy Cooperative (REC) considering heterogeneous prosumer preferences. 
The heterogeneity arises from prosumers' evaluation of \emph{energy contracts} through multiple societal and environmental criteria and the prosumers' private preferences over those criteria.
The prosumers engage in \emph{bilateral negotiations} with peers to mutually agree on periodical \emph{energy contracts/loans} that consist of an energy volume to be exchanged at that period and the return time of the exchanged energy. The prosumers keep an ordered \emph{preference profile} of possible \emph{energy contracts} by evaluating the contracts from their own valuations on the entailed criteria, and iteratively offer the peers contracts until an \emph{agreement} is formed. A prosumer embeds the valuations into a \emph{utility function} that further considers uncertainties imposed by demand and generation profiles. 
Empirical evaluation on real demand, generation and storage profiles illustrates that the proposed negotiation based strategy is able to increase the system efficiency (measured by \emph{utilitarian} social welfare) and fairness (measured by \emph{Nash} social welfare) over a baseline strategy and an individual flexibility control strategy.
We thus elicit system benefits from P2P flexibility exchange already with few agents and without central coordination, providing a simple yet flexible and effective paradigm that may complement existing markets.

% Development of an alternative energy market without inclusion of pricing. 
% Therefore, we present a bi-lateral negotiation strategy that seeks out the optimal energy contract considering the private preferences of prosumers. 
% Following is a simple case of finding optimal energy contract between prosumers for energy trading.
% In a prousmer marketplace

% and are reflected on their energy contracts with market operators. Integrating diverse and heterogeneous preferences as such into an exchange decision making problem is quite challenging.

\end{abstract}

% \begin{IEEEkeywords}
% Automated Negotiation, Energy Contract, Multiagent System, P2P energy exchange.
% \end{IEEEkeywords}

% Use this to place sponsorships
% \thanksto{Grid-Friends}

\section{Introduction}
\label{intro}
% The deregulated electricity markets refrain small-scale (residential) prosumers to actively participate in the wholesale market.
% The prosumers are instead serviced in retail markets, where they are individually metered by large suppliers~\cite{chao2001design}, through representative \emph{residential aggregators}.
% This scenario results in inefficiencies since the prosumers control and optimize their local energy usage individually without taking the overall demand and supply status into consideration. 
% The jointly coordinated prosumers' (distributed) energy resources potentially shape up the overall demand and offer significant value to the energy system by alleviating the need for investment in additional generation and transmission infrastructure~\cite{pudjianto2010value} and by reducing the fluctuations due to renewable power integration. 
% The prosumers in deregulated electricity market-era tend to individually control and optimize their local energy usage without taking the overall demand and supply scenarios into consideration as they are metered individually by large suppliers and served in retail market.
% ~\cite{chao2001design}. 
The joint coordination of prosumers' distributed energy resources (DER) in a Residential Energy Cooperative (REC) has the potential to shape the overall demand and to mitigate fluctuations caused by renewable integration.
However, properly incentivizing the prosumers to coordinate their locally owned distributed resources is quite a challenge, and justifiably a field of active research in Smart Grids. 
% Distributed optimization techniques facilitate the coordinations of the DERs that are owned and controlled by a single entity. However, these techniques are not readily applicable when these resources have different owners; at least not without considering the strategic interactions between the owners. 
% Efficiently designed liquid local energy markets, via centralized auction and profit sharing mechanism, 
Local energy exchange may offer incentives to the prosumers to engage in competition and in local trading~\cite{liu2017energysharing}. For energy communities, these mechanisms may need to take into account and balance several objectives, including, next to efficiency, altruism, or fairness of allocations~\cite{cornelusse2018optimal}. 
Prosumers in a REC may have diverse preferences over how their energy profiles are valued due to various societal and environmental factors. 
For instance, the prosumers may evaluate \emph{energy contracts} based on several criteria, e.g. \emph{self-sufficiency} or \emph{autarky}, \emph{cost of energy}, \emph{loss in flexibility}, \emph{sustainability}, and so on~\cite{baarslag2017}, resulting in a private valuation.
% Right now, market-based mechanisms are not fully designed to handle such heterogeneity in distributed decision making, where the prosumer-specific individual allocations are of absolute necessity. 
% Local market-based mechanisms are, however, not designed to handle such heterogeneity in distributed decision making, where the prosumer-specific individual allocations are of absolute necessity. 
While complete preferences would need to be computed and revealed for market based solutions, peer-to-peer (P2P) negotiation proceed iteratively, reducing the amount of information revealed.

Automated negotiation is an organic process of joint decision making where multiple stakeholders -- typically represented by autonomous agents -- with conflicting interests engage and make a decision~\cite{baarslag2014bid}. 
The negotiation approach contrasts market-based approaches, and its iterative nature provides a more natural model for low liquidity settings, in which personalized solutions need to be found in large outcome spaces. 
P2P negotiation within REC's is still a widely unexplored area of research, with a few exceptions; e.g., an automated negotiation protocol has been applied to address energy exchange between off-grid smart homes~\cite{alam2015}. 
However, their designed protocol imposes several key restrictions, in which only two \emph{exchange periods} over a day in which only equal amounts \emph{energy volume} can be exchanged. 

We present an \emph{automated negotiation approach} as an energy exchange mechanism to settle P2P energy contract as loans between prosumers in a REC. 
% The REC is represented as a Multiagent system, where the prosumers are mimicked by software agents.
During each negotiation session, a pair of prosumers (represented by software agents) engage in \emph{bilateral negotiation} by exchanging and eventually agreeing on \emph{energy contracts}, comprised of several \emph{negotiation issues} (here \emph{energy volume} and \emph{return time}). %Additionally, an energy exchange mechanism that is formed on the basis of borrowing and returning (of energy volume) essentially avoids the necessity of having a (financial) payment structure. 
The proxy agents evaluate \emph{offers} based on criteria that model the heterogeneous preferences of the users they represent: 1) \emph{loss in flexibility} (in local storage), and 2) \emph{autarky} or sustainability of the \emph{offers}. The agents are able to weigh these criteria differently, thereby enabling heterogeneity and trade-offs between the agents.
% Additionally, agents takes into account the uncertainty imposed by demand and renewable generation prediction while iteratively offering the \emph{energy contracts} with higher \emph{expected utility}.

The main contributions of this paper are as follows.
\begin{itemize}
  \item We propose a novel and intuitive \emph{negotiation} based strategy that considers \textit{heterogeneity} in prosumers through a \emph{distributed} and \emph{autonomous} agent model.%\footnote{We believe that this is the first work that considers heterogeneity in REC.}. 
  % \item Imposes robustness into the offered \emph{energy contracts} by considering uncertainty in demand and generation profiles while generating negotiation \emph{preference profiles}.
  \item We evaluate the performance of the proposed \emph{negotiation} based strategy over real residential demand, generation, and storage data to elucidate the efficiency of the strategy in increasing the social welfare over a baseline strategy and an individual flexibility control strategy.
\end{itemize}

The rest of the paper is organized as follows: Section~\ref{system_model} describes a residential prosumer model, and defines the \emph{energy contract} that is used in the negotiation process. Section~\ref{sec_negotiation} presents the negotiation based energy exchange strategy and the contextual notion of allocative efficiency. Simulation case studies are presented in Section~\ref{sec_sim}. Finally, Section~\ref{sec_conclusion} concludes the paper with a glimpse of possible follow-up research.

\section{Modeling Prosumers and Energy Contracts}
\label{system_model}
In this section, we present a prosumer by systematically modeling their load and generation profiles integrated with batteries. Later, we define \emph{energy contracts} with associated concepts, uncertainty management in planning, and aspects of the negotiation process. We assume that the energy cooperative forms a Microgrid and is located at the Low Voltage (LV) distribution network where the prosumers are physically connected to exchange energy.

\subsection{Prosumer Model}
\label{sec_modeling_prosumer}
A prosumer is assumed to be equipped with a renewable power generation unit (e.g. Solar Photovoltaic, PV) and a flexible resource (e.g. battery energy storage system). We represent a prosumer as a software agent, $i \in N$, where $N$ is the set of agents in the cooperative. Let the (predicted) generation profile (through PV panels) of an agent $i$ be represented by $\widetilde{Pv}_{i}(t),~\forall t\in T$, where $T$ is the set of time-periods. Similarly, the predicted load profile of the agent $i$, at $t$ is represented by $\widetilde{Ld}_{i}(t)$. In addition, the battery dispatch (load) profile is denoted as $Pb_{i}(t)$, and a choice of agent $i$. The battery state of charge is modeled by the following equation
\begin{align}
X_i(t) = X_i(t-1) + \eta_b \times Pb_i(t) \times \Delta t - \epsilon_b,
\end{align}
where, $X_i(t)$ is state of charge (SOC) of the battery at $t$ and is operated within a limit. The constant degradation of the battery is represented by $\epsilon_b$. The dispatched battery power, $Pb_{i}(t)$ is constrained to operate within a limit. The efficiency of the battery, $\eta_b$ is dependent on whether the battery is being charged (with efficiency $\eta_{b}^{c}$) or discharged (with efficiency $\eta_{b}^{d}$)
\begin{align}
\eta_{b} = \left\{\begin{matrix}
 \eta_{b}^{c}, & \text{if}~Pb_i(t) \geq 0\\ 
 \frac{1}{\eta_{b}^{d}}, & \text{otherwise}
\end{matrix}\right.
\end{align}
After self-consumption, the net demand of agent $i$ becomes
\begin{align}
\label{eq_net_demand}
\widetilde{Ld}_i^{net}(t) = \widetilde{Ld}_i(t)-\widetilde{Pv}_i(t).
\end{align}
An agent $i$ engages in a trade with a subset of peers $j\in J \subseteq N$ at $t$, and the volume of energy being traded with each other is denoted as $ex_{i,j}(t)$.
The residual of agent $i$ -- after self-consumption, followed by the (cumulative) exchange with the peers and the local battery activation -- is the energy either wasted or to be traded on the external market, and is presented by the following energy balancing equation.
\begin{align}
\widetilde{Ld}_i^{res}(t) = \widetilde{Ld}_i^{net}(t)+ \sum_{j\in J}ex_{i,j}(t) + Pb_i(t).
\label{eq_net_load}
\end{align}
% The above Eq.~\ref{eq_net_load} serves as the energy balancing equation.

\subsection{Energy Contracts: P2P energy lending}
\label{sec_contract_def}
% Negotiation in a form of energy exchange between prosumers without any financial payment.
A simple but effective contract to exchange flexibility in energy systems are \emph{energy loans}~\cite{Claessen2016}, which we here adapt to the P2P setting.
In automated multi-issue negotiation, agents negotiate over several \emph{issues} with a target to achieve an agreement -- a value attribution to those issues -- that generates a socially optimal outcome for the participating agents. 
% Let $\Omega=\left { I_1, I_2, \cdots, I_M\right }$ represents the set of all possible outcomes of the negotiation. 
We consider energy loans parameterized by two important \emph{issues} over which the agents negotiate:
\begin{itemize}
\item[1.] The volume of energy to be traded between two agents (denoted by $q \in \mathcal{Q} \subseteq \mathbb{R}$, where $\mathcal{Q}$ is a discrete set of energy volumes).
\item[2.] The time of receiving the energy back (denoted by $\tau \in \mathcal{T} \subseteq \bm{Z^{+}}_{>0}$, where $\mathcal{T}$ is a discrete set of positive time periods).
\end{itemize}
A \emph{negotiation domain} $\Omega$ comprises all possible \emph{energy contracts}, i.e. $\Omega=\mathcal{Q}\times\mathcal{T}$. Every $\omega = (q,\tau) \in \Omega$ is a potential \emph{energy contract} (or loan) within the multi-issue negotiation that specifies a value for each \emph{issue}.

The energy volume $q$ and the return time $t+\tau$ influence respective energy profiles for both negotiating agents and consequently affect their local flexibility dispatch. 
% Therefore, different \emph{energy contract} (i.e. different values of $I:=\{q,\tau\}$) yield different energy profiles. 
% As the agents negotiate over an \emph{energy contract}, an \emph{offer} is also a form of an \emph{energy contract} that the agents repeatedly exchange within a particular negotiation round. 
% Let $I_{i\rightarrow j}^{r, t}$ represent an offer made by agent $i$ to agent $j$ during a negotiation round $r$ for the settlement period $t$. 
% Finally, the agreement between 
% Next, we define the strategy agent $i$ undertakes to evaluate an \emph{energy contract} or \emph{offer}, $I$. 
As depicted before, agents may have varied preferences over a predefined set of criteria, i.e. the agents tend to weigh the criteria differently. Let $\mathbb{C}$ represent the set of criteria upon which the agents state their preferences. In this paper, we assume an agent evaluates an energy profile -- resulting from an \emph{energy contract} -- based on two criteria: \emph{loss in flexibility} and \emph{autarky} i.e. $\mathbb{C}:=\left\{ c_1=loss~in~flexibility, c_2=autarky \right\}$. 
The weight an agent $i$ places on criterion $c\in \mathbb{C}$ can be represented as a scaler $\lambda_{c}^{i}$ (where $\lambda_{c}^{i}$ are normalized weights, i.e. $\sum{_c}\lambda_{c}^{i}=1$) and is private to $i$. We assume, the weights are known to agent.
An \emph{evaluation function}, $e_{c,t}^{i}(\omega)$ is defined that denotes how an \emph{energy contract} performs, at $t$, from the perspective of criterion $c$ given the private preferences of agent $i$. 
Additionally, an agent maintains a planning horizon, $w$ that represents how far ahead of the agent looks while deciding about the \emph{contracts}. The planning horizon depends on the uncertainty on the demand/generation prediction.

\textit{Criterion 1:} Criterion \emph{loss in flexibility} measures the emergent loss (in energy) due to the round-trip efficiency of the flexibility (e.g. battery) dispatch resulting from implementing an \emph{energy contract}. 
% In other words, agent $i$ measures the quality of an \emph{energy contract} by quantifying \emph{how much} the contract reduces the loss (presented by an \emph{evaluation function}) scaled by \emph{how} the agent values (or prefer) \emph{loss in flexibility} criterion (presented by the associated weight).
The \emph{evaluation function} associated with \emph{loss in flexibility} is defined as
\begin{align}
e_{c_1,t}^{i}(\omega)=\sum_{k=t}^{k=t+w}Pb_i(k) + \Theta(\bf{X_{i}}),
\end{align}
where $\Theta(\bf{X_{i}})$ is the offset power required to adjust the resulting SOC. Therefore, $\omega$ directly influences the battery dispatch power $Pb_i(t)$ through the energy balancing equation, i.e. Eq.~\ref{eq_net_load}.

\textit{Criterion 2:} Autarky in an \emph{energy contract} signifies the sustainability of the contract, which actually measures the total (estimated) energy to be traded on the external market provided that the \emph{energy contract} is implemented.
% An agent measures the quality of an \emph{energy contract} by quantifying \emph{how much} the contract increases the \emph{autarky} scaled by \emph{how} the agent values \emph{autarky}. 
The \emph{evaluation function} associated \emph{autarky} is formally defined as
\begin{align}
e_{c_2,t}^{i}(\omega)=\sum_{k=t}^{k=t+w}|\widetilde{Ld}_i^{res}(k)|.
\end{align}
Agent aggregates the weighted \emph{evaluation function} of individual criterion to measure the quality of an \emph{energy contract}. The \emph{utility} function is defined as
\begin{align}
f_{i,t}(\omega)= \sum_{c}\lambda_{c}^{i} e_{c,t}^{i}(\omega).
\label{eq_utility}
\end{align}

\subsection{Dealing with Uncertainty}
The load profile $\widetilde{Ld}_{i}(t)$ and generation profile $\widetilde{Pv}_{i}(t)$ of an agent $i$ are predicted signals and are  potential sources of uncertainties. 
The \emph{utility} function defined in Eq.~\ref{eq_utility} is, therefore, unable to provide robust scheduling of local flexibilities. 
% The \emph{utility} function facilitates determining the \emph{preference profile}, which is an important step an agent performs in the negotiation process. 
We utilize a set of stochastic scenarios of predicted net load profiles $\widetilde{Ld}_{i}^{net}$ and calculate the \emph{expected utility} of an \emph{energy contract}~\cite{chakra2016sg}. The scenarios of predicted net load profile are generated by taking samples from a \emph{Gaussian Process} comprising of 1) \emph{Gaussian} error Probability Density Functions (PDF), for each of the discrete lags $l$ in planning horizon $w$, i.e. $l=1,\cdots,w$, and 2) a \emph{Gaussian} PDF that models the interdependency between net load of two consecutive periods. 
The predicted net load for scenario $s\in \mathcal{S}$ is then determined as $\widetilde{Ld}_{i}^{net}(t+l|t,s)=\widetilde{Ld}_{i}^{net}(t+l|t) + d_{i}(l,s)$, where $d_{i}(l,s)$ is sampled from the aforementioned \emph{Gaussian Process} and $\widetilde{Ld}_{i}^{net}(t+l|t)$ is the predicted net 
demand for period $t+l$ when predicted at $t$\footnote{Pseudo-predictions of the net demand are generated by adding \emph{Gaussian} noise to the real net demand signal.}. Now, we can define the \emph{expected utility} an \emph{energy contract} could provide by
\begin{align}
% \mathbb{E}[f_{i,t}(I)]= \sum_{s\in \mathcal{S}}Pr(s) \cdot \times\sum_{c}\lambda_{c}^{i} f_{c,t}^{i}(I, s),
\mathbb{E}[f_{i,t}(\omega)]= \sum_{s\in \mathcal{S}}Pr(s) \cdot f_{i,t}(\omega,s),
\label{eq_ex_utility}
\end{align}
where $Pr(s)$ is the probability of the scenario $s$ and $f_{i,t}(\omega, s)$ is the modified \emph{utility} of an \emph{offer} $\omega$ considering 
the net predicted load scenario $s$. We assume the scenarios are equiprobable, and thus $Pr(s)=1/|\mathcal{S}|$. Therefore, in the \emph{negotiation process}, the agents distributively search through the $\Omega$ to jointly agree on an \emph{energy contract} that maximizes their perspective \emph{expected utilities}.

% -- of an agent 
% measure the preference the quality of an \emph{energy contract}
% The strategy space generation requires predicted load profile for the window of $w$, $\widetilde{Ld_i^{net}}(t)$ to $\widetilde{Ld_i^{net}}(t+w)$. Instead of using a single point prediction, the agent generates a number of prediction scenarios and calculates associated \emph{expected} utility of a strategy. 

% The negotiation goes in rounds to seek for the agreement (as the final contract) that maximizes the joint \emph{utility} of the agents.

% \begin{figure}[h]
% \centering
% \includegraphics[scale=0.205]{fig_contract_example_2.png}
% \caption{Contract realization between Agent $a$ and $b$. [WILL REPLACE FIGURE 1 WITH DESCRIPTION]}
% \label{fig_contract}
% \end{figure}

% \section{Energy Contracts}

\section{A Negotiation based Exchange Mechanism}
\label{sec_negotiation}
Agents engage in a \emph{bilateral negotiation} to seek for an agreement on an \emph{energy contract}. 
Given the residential energy cooperative settings of several connected prosumers, the process may be understood as a \emph{multilateral negotiation}, emerging from multiple bilateral P2P pairwise negotiations\footnote{An alternative approach could be \emph{multi to multi negotiation}. We do not entertain that option since \emph{multi to multi negotiation} typically require a mediator or a centralized coordination, as per the current state of the research.}. As the \emph{negotiation protocol}, we implement the \emph{alternating offers protocol}~\cite{baarslag2014bid}, which is commonly used in automated multi-issue negotiation settings.
We assume that each agent is limited to interact once with one other agent in each particular time period.
% TODO: This requires clarification
% \todo{Re-frame as multilateral negotiation, but \textit{through} P2P interaction. Multi to multi negotiation currently often depends on a mediator or centralized coordination.}
Several important aspects of the negotiation process are detailed in the following. 
% Refer to~\cite{baarslag2014bid} for the detailed descriptions.

\textit{Agreement:} An \emph{agreement} is an \emph{energy contract} that is approved by both negotiating agents, and can be denoted by $\omega^{*}=(q^{*},\tau^{*})$.

\textit{Reservation value:} The private value a negotiating agent keeps as a criterion to accept an offer. Additionally, it also represents the outside option in case of a disagreement. In this paper, an agent $i$ sets the \emph{reservation value} as a \emph{quantile} of the distribution of $\mathbb{E}[f_{i,t}(\omega)],~\forall\omega\in\Omega$ for negotiating at $t$. Additionally, the \emph{energy contracts} that generate \emph{expected utility} higher than the \emph{reservation value} forms a so-called \emph{aspiration region}.

% The \emph{reservation value} essentially defines the \emph{degree of cooperativeness} of an agent; higher quantile represents lower cooperation and vice-versa.
\textit{Deadline:} The maximum number of rounds of a negotiation before which the agents should reach an outcome. If no \emph{agreement} is formed after the \emph{deadline}, the negotiation fails.
% Alternative, one of the agents may adjust her \emph{reservation value} and restart the negotiation session.\\

\textit{Preference profile:} A negotiating agent contains a \emph{preference profile} that accumulates an ordered set of the issues in the \emph{negotiation domain}. The agent creates such a profile by ordering the issues according to their \emph{expected utility}, defined in Eq.~\ref{eq_ex_utility}. 

\textit{Fair outcome:} An important measure to quantify the fairness in an \emph{outcome} could be conducted by 
determining the \emph{Nash solution}. The \emph{Nash solution} is essentially the outcome that maximizes the product of the utilities (Eq.~\ref{eq_ex_utility}), achieved from an \emph{energy contract}, of negotiating agents (e.g. agent $i$ and $j$).
\begin{align}
\omega_{\rm{Nash}}(t)=\underset{\omega\in\Omega}{\max}~\mathbb{E}[f_{i,t}(\omega)]\cdot \mathbb{E}[f_{j,t}(\omega)].
\label{eq_nash_solution}
\end{align}

\begin{figure}[h]
\centering
\includegraphics[scale=0.35]{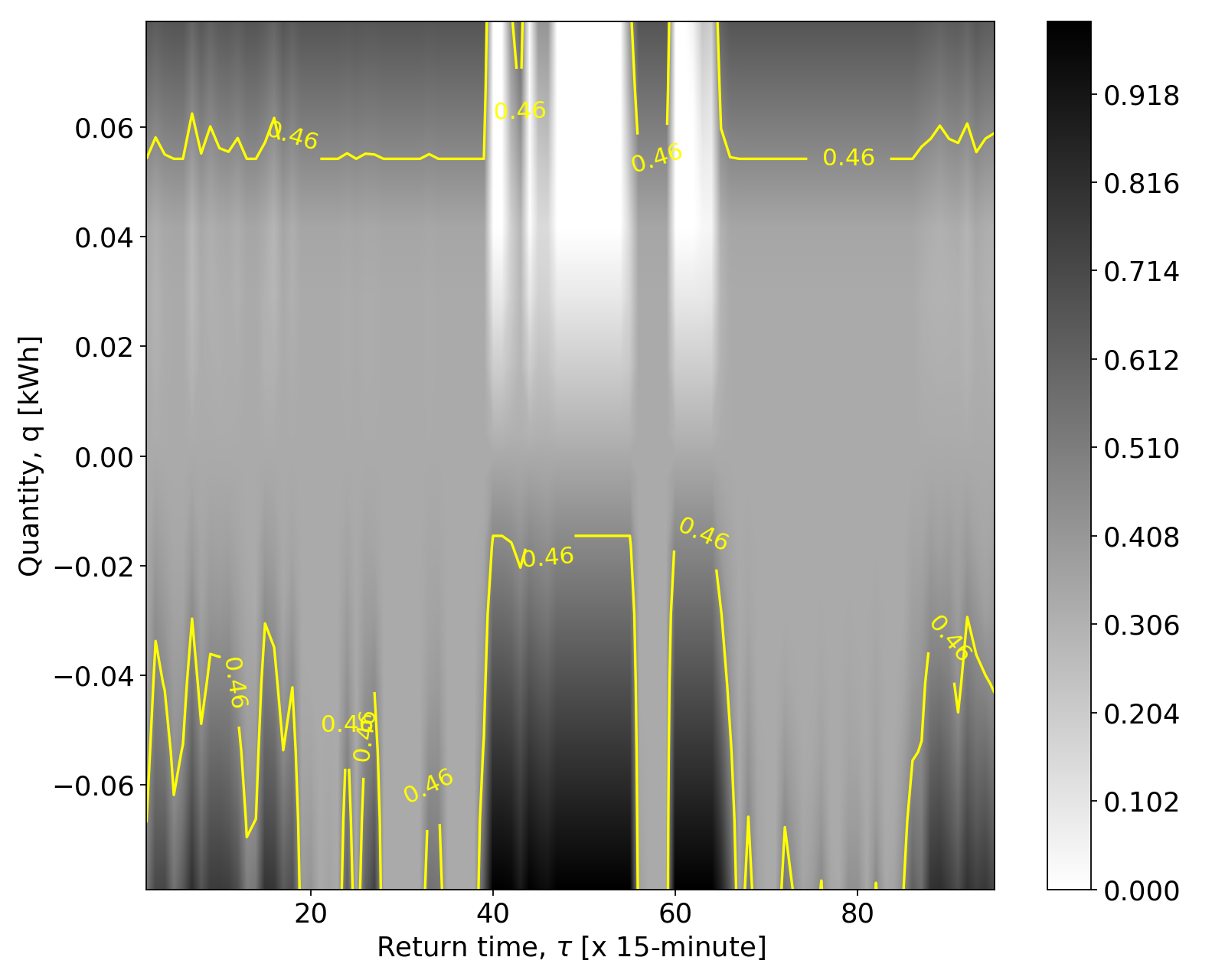}
\caption{An exemplary \emph{negotiation domain}. The \emph{issues} are presented at the axes whereas the \emph{expected utility} of the contracts is plotted as a heat-map. The contoured regions are the \emph{aspiration region} as they contain the preferred \emph{energy contracts} with the \emph{expected utility} higher than the \emph{reservation value}.}
\label{fig_strategy_space}
\end{figure}

Figure~\ref{fig_strategy_space} shows an exemplary \emph{negotiation domain} and the associated \emph{expected utility} (normalized within unit-range) of all possible \emph{energy contracts} calculated from the perspective of an agent that has a \emph{reservation value} quantile of 70\% of the distribution that results in a value of $\textrm{0.46}$ in the normalized scale. The demarcation of the \emph{aspiration regions} in the \emph{negotiation domain} are outlined through the contoured line. The figure illustrates that expected utility is smooth over quantity and nonlinear over return-time.

% \comments{Improve 3d plot}{Make into 2d. $\tau$ = return time horizontally, utility vertical, indicate quantity with blobs (or vice versa).}
\begin{algorithm}[h]
\scriptsize
\DontPrintSemicolon
\Begin{
$\Omega \longleftarrow \textrm{createNegotiationDomain}(\mathcal{Q}, \mathcal{T})$\;
$A.\textrm{generateOrderedPreferenceProfile}(\Omega, t)$\;
$B.\textrm{generateOrderedPreferenceProfile}(\Omega, t)$\;
$r \longleftarrow 0$\;
$offerAccepted \longleftarrow False$\;
\While{$r < deadline$~\rm{AND}~NOT~$offerAccepted$}{
  \eIf{$r\%2=0$}{
      $offer \longleftarrow A.\textrm{makeOffer}(t)$\;
        $offerAccepted \longleftarrow B.\textrm{acceptOffer}(offer, t)$\;
    }
    {
      $offer \longleftarrow B.\textrm{makeOffer}(t)$\;
      $offerAccepted \longleftarrow A.\textrm{acceptOffer}(offer, t)$\;
    }
    $r\longleftarrow r+1$\;
}
\eIf{$offerAccepted$}{
  $A.\textrm{implementContract}(offer, t, B)$\;
  $B.\textrm{implementContract}(offer, t, A)$\;
}{
  $A.\textrm{implementReservePlan}(t)$\;
  $B.\textrm{implementReservePlan}(t)$\;
}

return $offerAccepted$\;
}
\caption{NEGOTIATE-Contract($A$, $B$, $t$, $\mathcal{Q}$, $\mathcal{T}$)\label{al_negotiation}}
\end{algorithm}
\normalsize
Algorithm~\ref{al_negotiation} describes the high-level algorithm of the negotiation process between two agents $A$, and $B$ at time $t\in T$.
% \footnote{We present a simplified version of the algorithm, and it can easily be implemented in a Multiagent environment.}. 
The process starts with creating negotiation domain $\Omega$ that will be used by both agents. 
Agents then generate perspective ordered \emph{preference profiles} by evaluating all possible contracts in $\Omega$ while considering their \emph{expected utility} over a planning horizon $w$. Subsequently, an \emph{alternating offers protocol} is implemented where, in each round, 
one of the agents proposes 
an offer (picked from the ordered \emph{preference profile}) to the other agent until an \emph{agreement} is reached or the \emph{deadline} is encountered. 
In case an \emph{agreement} is reached, the agents implement the agreed \emph{energy contract}. Otherwise, the plans associated with the \emph{
reservation values} are implemented by each agent.
While implementing an \emph{energy contract}, an agent (for instance, $A$) amends to an existing exchange pool by stating \emph{how much} energy ($q^{*}$) 
to be traded with \emph{whom} (for instance, $B$) and \emph{when} ($t$) as well as by listing the \emph{same} volume of energy ($-q^{*}$) is committed to 
be traded back at $(t+\tau^{*})$ from $B$.

\subsection{Efficiency and Fairness}
In this section, we define the following strategies -- apart from the proposed negotiation based energy exchange -- to illustrate the efficiency of the proposed strategy. 
% , compare their outcomes, and draw insights regarding the allocative efficiency.
\begin{itemize}
\item \textit{No flexibility, $s_0$:} The prosumers do not activate their batteries and only trade residuals with external market.
\item \textit{Individual control, $s_1$:} This strategy is being currently utilized in the real residential setting, where the prosumers activate their local batteries, individually control the batteries and trade the residuals with external market. However, prosumers do not engage in trading with each other.
\item \textit{Negotiation and control, $s_2$:} The proposed strategy where prosumers engage in bilateral negotiation over \emph{energy contracts} with peers, implement the \emph{agreement}, and finally activate their batteries to control the residual energy. 
The remaining energy is traded in external market.
\end{itemize}
The properties of the strategies are briefed in Table~\ref{tbl_scn}.
\begin{table}[h]
\centering
\caption{Properties of the strategies.}
\label{tbl_scn}
\begin{tabular}{|r|r|r|r|}
\hline
Strategy & Local trading & Flexibility activation\\ \hline
\emph{No flexibility} & No & No \\ 
\emph{Individual control} & No & Yes  \\ 
\emph{Negotiation and control} & Yes & Yes \\ \hline
\end{tabular}
\end{table}
Now, we define the \emph{utility} of an agent achieved by applying a particular strategy. Note that, it differs from the \emph{utility function} defined in Eq.~\ref{eq_utility} which measured the quality of an \emph{energy contract}.
% The proposed \emph{Negotiation and control} strategy follows a closed-loop methodology. 
% Let $I_{t}^{*}$ represents the \emph{agreement} achieved at time $t$, and the agreement history is represented by $\mathcal{I}=\{I_{t}^{*}\}_{t}^{T}$. 
The \emph{utility} of the proposed strategy $s_2$ considers the realized energy profile, after periodically negotiating and implementing the \emph{agreement}. The \emph{utility} is, therefore, similar to Eq.~\ref{eq_utility} but taking into account the realized energy profile and the consequent battery dispatch.

\begin{align}
u_{i}(s_2) = \lambda_{1}\times \left [ \sum_{t}^{T}Pb_i(t) + \Theta(\bf{X_{i}})\right] + \lambda_{2}\times \left [ \sum_{t}^{T}\left |Ld_i^{res}(t)\right | \right].
\end{align}
For strategy $s_{0}$, the $u_{i}(s_0)$ only considers the \emph{autarky} components (without the energy exchange component, i.e. $\sum_{j\in J}ex_{i,j}(t)$ in Eq.~\ref{eq_net_load}). And for strategy $s_{1}$, the $u_{i}(s_1)$ considers both criteria, but again without the flexibility component.
In order to validate the efficiency of the strategies $\mathbb{S}:=\{s_0, s_1, s_2\}$ in improving the social welfare, we define the \emph{utilitarian social welfare} as $sw_{s} = \sum_i^{N} u_i(s)$
% \begin{align}
% sw_{s} = \sum_i^{N} u_i(s),
% \label{eq_sw}
% \end{align}
for all $s\in \mathbb{S}$. Moreover, we quantify the relative fairness of a strategy $s$ (to another strategy $h$) based on the \emph{Nash social welfare} criterion, an established concept of fairness~\cite{moulin2004fair}, as following
\begin{align}
nw_{s|h} = \prod_i^{N} \left ( u_i(s) - u_i(h)\right ).
\label{eq_fairness}
\end{align}
% where $s \neq k$.
% An important measure to quantify the fairness in an \emph{outcome} could be conducted by 
% determining the \emph{Nash solution}. The \emph{Nash solution} is essentially the outcome that maximizes the product of the utilities (Eq.~\ref{eq_ex_utility}), achieved from an \emph{energy contract}, of negotiating agents (e.g. agent $i$ and $j$).
% \begin{align}
% \omega_{\rm{Nash}}(t)=\underset{\omega\in\Omega}{\max}~\mathbb{E}[f_{i,t}(\omega)]\cdot \mathbb{E}[f_{j,t}(\omega)].
% \label{eq_nash_solution}
% \end{align}

\section{Numerical Simulation and Discussion}
\label{sec_sim}
In this section, we consider two cases of varied scaled cooperatives to empirically evaluate different aspects of the proposed strategy. 
\begin{itemize}
\item Case 1: \textit{Cooperative of 2 agents} presents the effects of the proposed strategy on the residual demand and consequent battery dispatch, and the agents' \emph{negotiation} domain exploring phenomena.
\item Case 2: \textit{Cooperative of 9 agents} verifies the quality of the allocation achieved by the proposed strategy from the perspectives of efficiency and fairness.
\end{itemize} 
The aforementioned cases assume the local flexibility (i.e. battery) is owned privately and controlled individually by the prosumers\footnote{The total simulation period is taken as 20 days with 15-minute of granularity, i.e. $\Delta t=15$. The planning horizon $w$ is set out to be 48-hours. The number of scenarios $|\mathcal{S}|$ is set to 100. The set $\mathcal{Q}$ contains 10 discrete energy quantities, and the set $\mathcal{T}$ contains discrete time steps of $\{2, 3, \cdots, w\times\Delta t\}$. The \emph{deadline} of a negotiation session is set out to be $\textrm{5000}$ rounds.}. 
% We utilize the real residential energy and battery power profiles collected from project collaborator\footnote{The total simulation period is taken as 20 days with 15-minute of granularity, i.e. $\Delta t=15$. The planning horizon $w$ is set out to be 48-hours. The number of scenarios $|\mathcal{S}|$ is set to 100. The set $\mathcal{Q}$ contains 10 discrete energy quantities, and the set $\mathcal{T}$ contains discrete time steps of $\{2, 3, \cdots, w\times\Delta t\}$. The \emph{deadline} of a negotiation session is set out to be $\textrm{5000}$ rounds.}.

\begin{figure*}[h] 
    \centering
  \subfloat[Execution of an \emph{agreement}.]{%
       \includegraphics[width=0.40\linewidth]{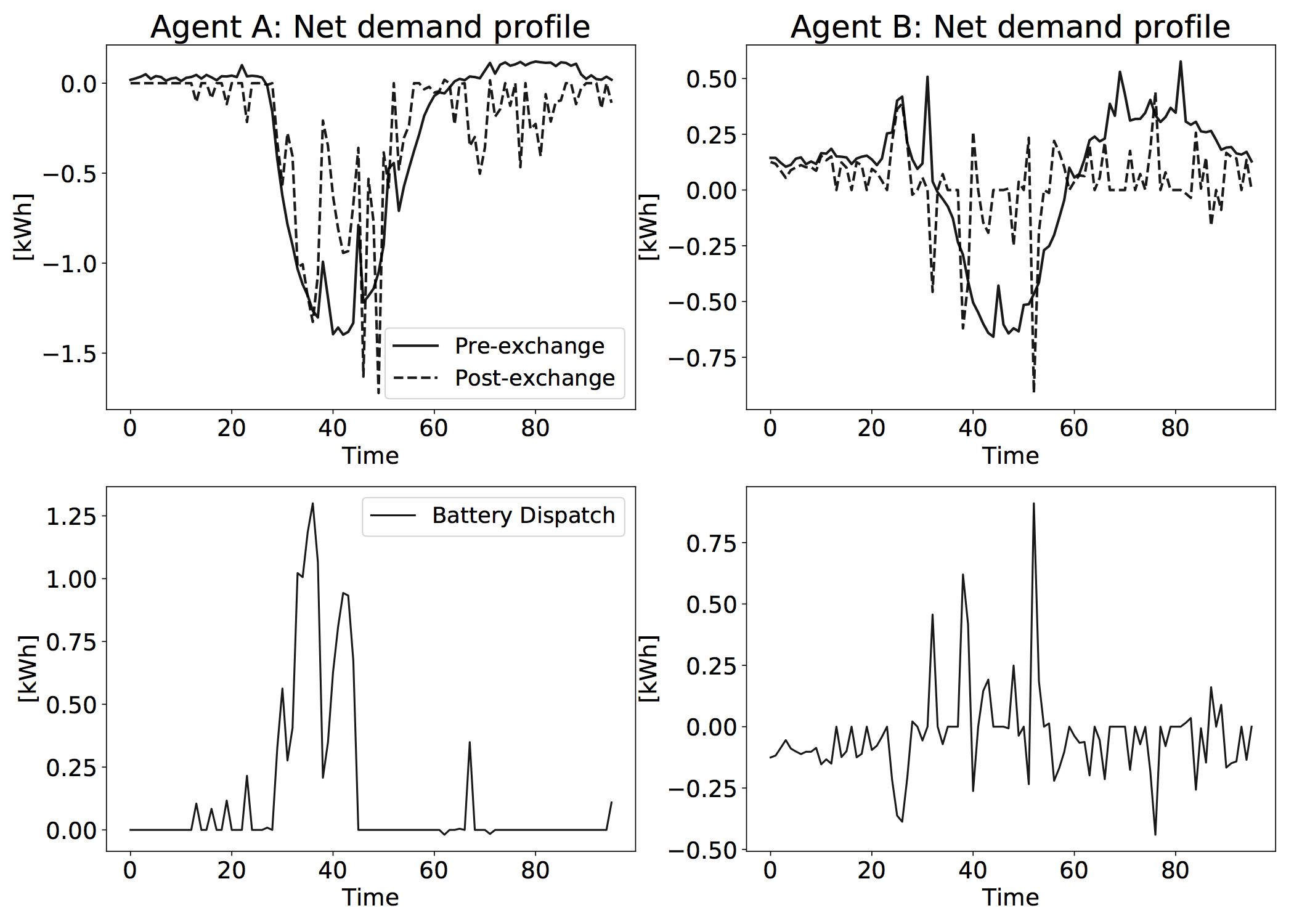}}
      % \label{fig_outcome_space}
      \hfill
  \subfloat[Outcome space: \emph{Agreement} at round 792.]{%
        \includegraphics[width=0.32\linewidth]{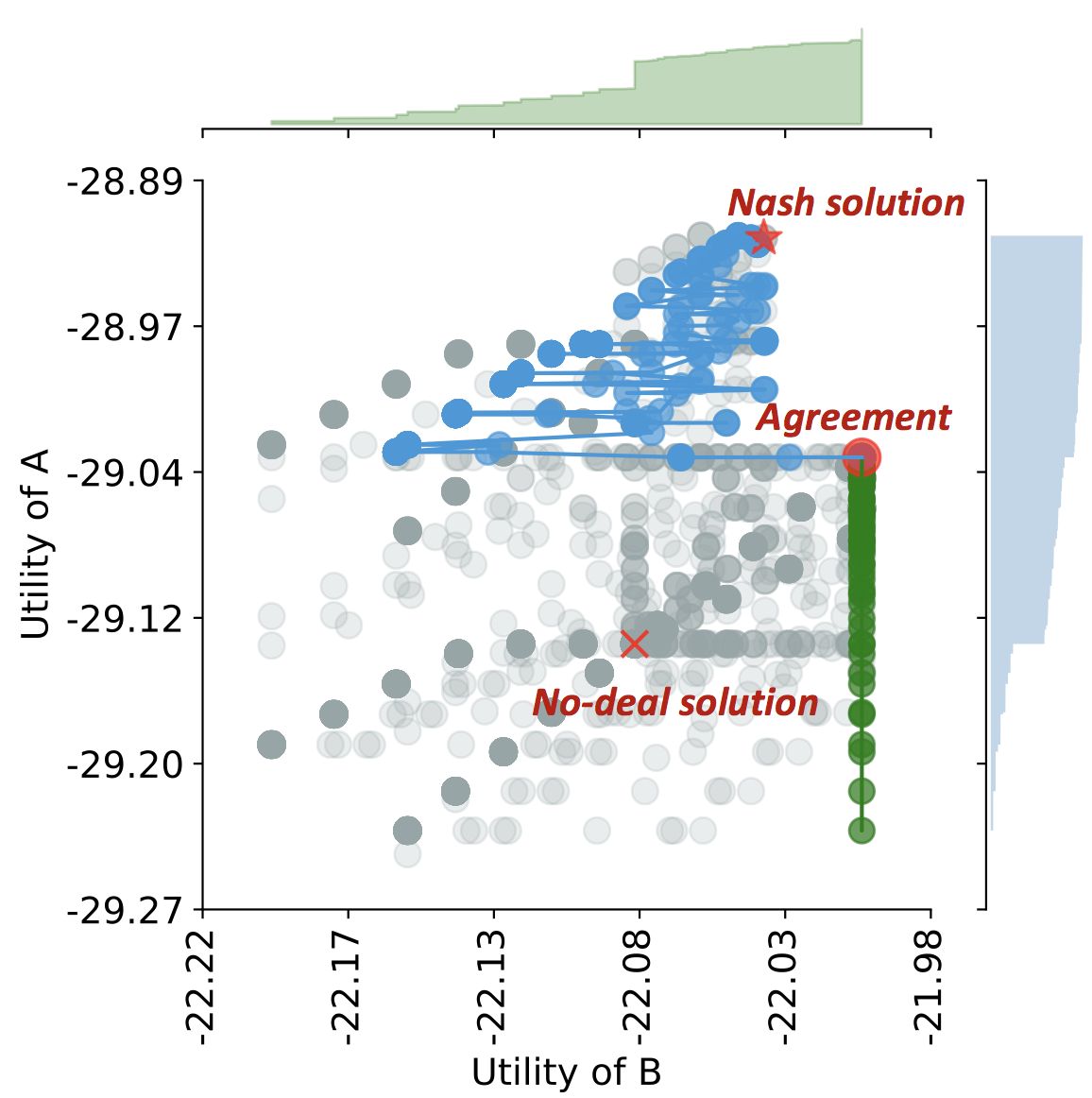}}
    % \label{fig_outcome_space_contour}
    \hfill
  \subfloat[Issue space.]{%
        \includegraphics[width=0.28\linewidth]{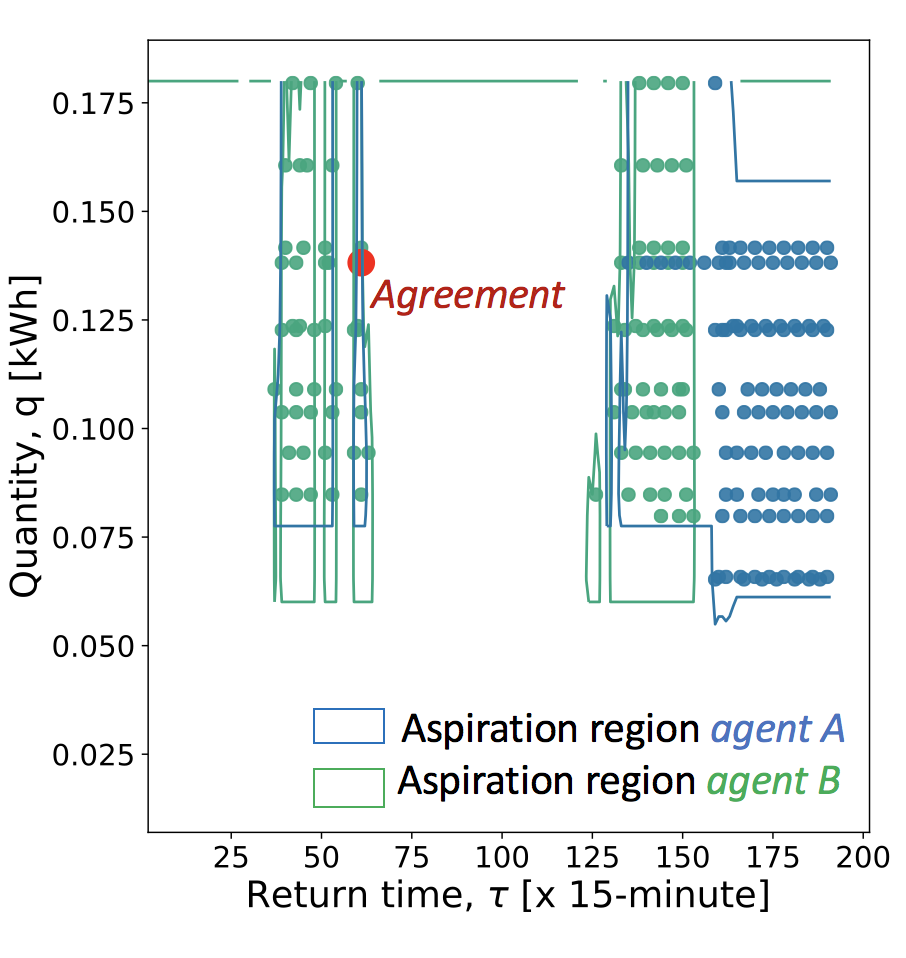}}
    % \label{fig_nash_distace}
  \caption{(a) Effects of executing \emph{agreements} -- on net demand and battery dispatch profiles -- reached through negotiation between two agents. (b) An exemplary outcome space of \emph{agent A} and \emph{agent B} with marginal cumulative distributions of their \emph{expected utilities}. The \emph{Nash solution}, \emph{agreement} and \emph{No-deal solution} are plotted to illustrate the relative distances. (c) The \emph{negotiation domain} with the juxtaposed \emph{aspiration regions} bounded by contour lines for both agents. The negotiation traces, represented by the colored scatters, depict the behavior of the agents with the opponents as both of them converge to the \emph{agreement}.}
  \label{fig_quality} 
\end{figure*}

\subsection{Case 1: Cooperative of 2 Agents}
% We present the \emph{2 Agents} case to elucidate the effects of negotiation based energy exchange between two agents on their energy profiles and resultant battery dispatch. 
Flexibility activation through battery enhances the potential benefits as two agents could negotiate even when their net demand status are equal (i.e. both positive or negative).
The specification of the agents with associated battery information is provided in Table~\ref{tbl_agent_spec}. The charging and discharging rates of these batteries are 1.3kW and 3.3kW, respectively. The SOC the batteries are operated within 20\% to 90\% of the respective capacity, 
and the degradation rate is set as 0.4\% of the same. As pointed in the Table, \emph{agent A} values criterion $c_2$ than criterion $c_1$; that is the agent places higher preferences on \emph{autarky}, while \emph{agent B} prefers \emph{loss in flexibility} more.

\begin{table}[t]
\centering
\caption{Agent Specifications.}
\label{tbl_agent_spec}
\begin{tabular}{|r|r|r|r|r|r|}
\hline
Agent & Reservation (\%) & $\lambda_{c_1}$ & $\lambda_{c_2}$ & Capacity[kWh] & Efficiency \\ \hline
\bf{A} & 52 & 0.33 & 0.67 & 6.8 & 0.9 \\ 
\bf{B} & 50 & 0.71 & 0.29 & 7.0 & 0.8 \\  \hline
\end{tabular}
\end{table}

% \begin{figure}[h]
% \centering
% \includegraphics[scale=0.25]{fig_two_agents.pdf}
% \caption{Effects of executing \emph{agreements} -- on net demand and battery dispatch profiles -- reached through negotiation between two agents.}
% \label{fig_two_prosumer}
% \end{figure}
Figure~\ref{fig_quality}(a) shows effects of energy exchange (through negotiation) and resulting battery dispatch between \emph{agent A} and \emph{agent B}. The residual demand profiles resulting from negotiation clearly reflect the preferences of the agents. For instance, the battery dispatch profile of \emph{agent B}, who cares more about the \emph{loss in flexibility}, exhibits a relative fluctuating signal that results in an almost neutralized losses. 
% Additionally, due to the lower efficiency of the battery (of \emph{agent B}) coupled with the degradation rate, the agent tends to shorten the delay in battery power dispatch.
The apparent fluctuations in the battery dispatches are due to the fact that they both implement a naive battery scheduling technique, as described in Section~\ref{sec_modeling_prosumer}. However, in the proposed framework, agents can easily mitigate such fluctuations by integrating an additional cost function (that penalizes such behavior) into their utility 
function, and placing a higher weight on that cost function.

Now, we analyze the exploration of \emph{negotiation domain} by agents while reaching an \emph{agreement}. The battery specifications of the agents are kept identical and similar to \emph{agent A}. The \emph{reservation quantile} are kept as 95\% for the agents. 
% \begin{figure}[h]
% \centering
% \includegraphics[scale=0.32]{fig_outcome_space_v5.png}
% \caption{An exemplary outcome space of \emph{agent A} and \emph{agent B} with marginal cumulative distributions of their \emph{expected utilities}. The \emph{Nash solution}, \emph{agreement} and \emph{No-deal solution} are plotted to illustrate the relative distances.}
% \label{fig_outcome_space}
% \end{figure}
% \begin{figure}[h]
% \centering
% \includegraphics[scale=0.32]{fig_contour_outcome.pdf}
% \caption{The \emph{negotiation domain} with utility distribution of \emph{agent B} w.r.t. Figure~\ref{fig_outcome_space}. The negotiation traces depict the behavior of the local \emph{agent B} with opponent \emph{agent A} (and vice-versa) while converging to an \emph{agreement}.}
% \label{fig_outcome_space_contour}
% \end{figure}
% \subsubsection{Quality of the outcome}
Figure~\ref{fig_quality}(b) depicts a two-dimensional outcome space that emerges from the negotiation interactions between the agents, and their marginal cumulative distributions of (expected) \emph{utility} over \emph{negotiation domain}. Noticeably, while the \emph{agreement} generates the highest \emph{utility} for \emph{agent B}, \emph{agent A} needs to compromise to reach the \emph{agreement}. Although, the \emph{agreement} does not reach the \emph{Nash solution}, it still yields \emph{utilities} that are located over 95\% quantile range of the distributions. The \emph{no-deal solution} defines the situation when the agents do not engage in negotiation, and consequently do not exchange energy.
The trace of negotiation -- from the perspective of individual agents -- illustrate the power of heterogeneous preferences and the multi-issue setting, because the agents are able to explore their \emph{iso-utility} curves, and concede until an \emph{agreement} is found. The corresponding \emph{negotiation domain}, with the issues and \emph{aspiration regions} of both the agents are illustrated in Figure~\ref{fig_quality}(c). The \emph{agreement} is located at one of the intersections of the agents' contoured region of \emph{aspiration value}.
% The traces elucidate that \emph{agent B} proposes offers that are situated in the higher \emph{utility} regions whereas \emph{agent A} starts by offering from the region that incurs lower \emph{utility} (from the perspective of \emph{agent B}), which actually located in the higher \emph{utility} region of \emph{agent A}.

% \begin{figure}[t]
% \centering
% \includegraphics[scale=0.53]{fig_residual_demand_agent_41.pdf}
% \caption{Residual demand for agent 41.}
% \label{fig_residual_demand_412}
% \end{figure}
\begin{figure}[h]
\centering
\includegraphics[scale=0.4]{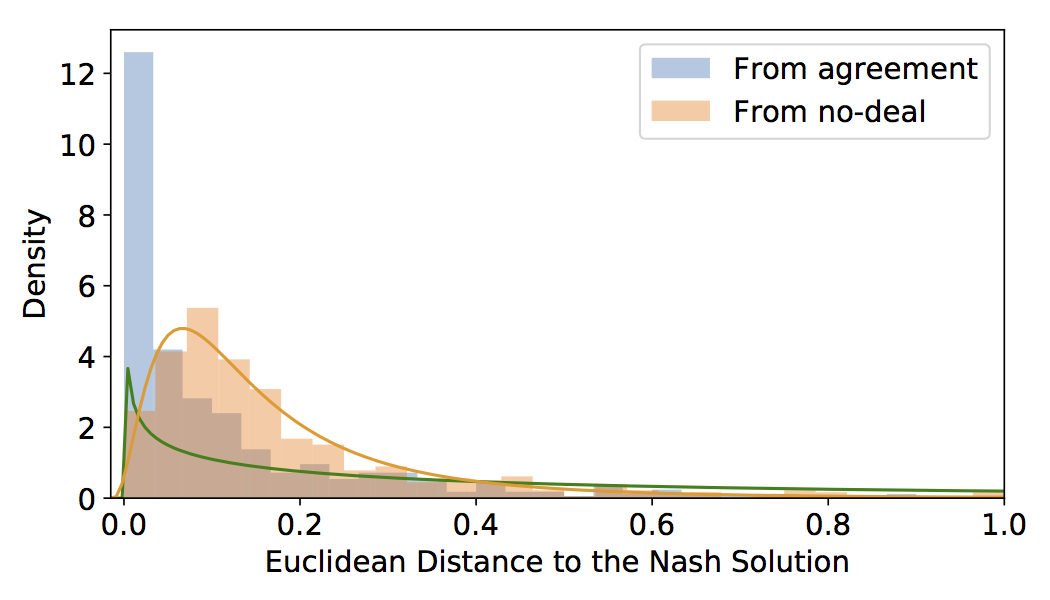}
\caption{Distributions of the 
Euclidean distance (fitted via \emph{F}-distributions) to the \emph{Nash solution} from the \emph{agreements} and the \emph{no-deal solutions} in the \emph{utility/outcome space}. The resulting \emph{agreements} are more inclined to be the \emph{Nash solutions}, hence confirm the fairness of the energy allocations.}
\label{fig_nash_distace}
\end{figure}

% \begin{figure}[h]
% \centering
% \includegraphics[scale=0.25]{fig_contour_outcome.png}
% \caption{Alternative of Figure~\ref{fig_quality}(c).}
% % \label{fig_nash_distace}
% \end{figure}

% \todo{We are interested in social welfare and how close we got to the optimal outcome... Could we have done better? (As an estimate/proxy, we use (normalized within unit range?) Euclidean distance)}
\subsection{Case 2: Cooperative of 9 Agents}
\begin{figure}[h]
\centering
\includegraphics[scale=0.40]{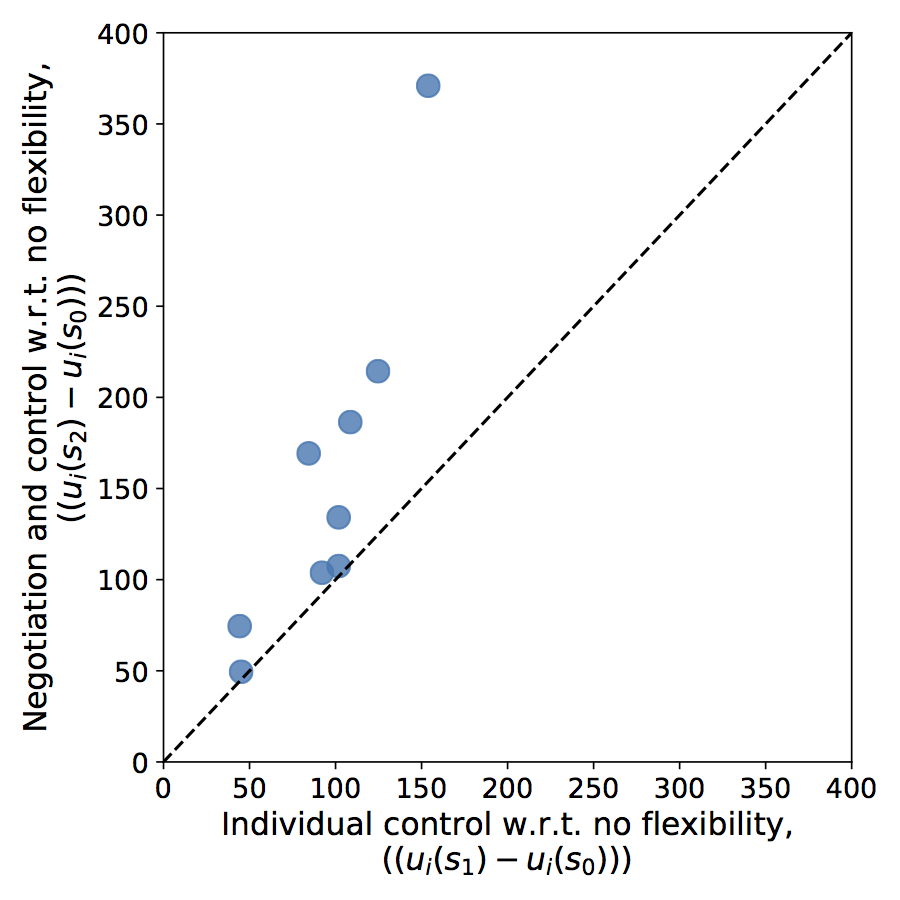}
\caption{Agents utility improvement of \emph{individual control} strategy (horizontal) and \emph{negotiation and control} strategy (vertical) over \emph{no flexibility} baseline. Agents are above the dashed equal improvements line, hence our newly proposed strategy dominates \emph{individual control}, while also improving relative fairness ($nw_{s_2|s_0} \approx 1.38 \cdot 10^{19} > nw_{s_1|s_0} \approx 3.35 \cdot 10^{17}$).}
\label{fig_improvement_sw}
\end{figure}

In this case, we analyze a higher scaled cooperative with 9 agents comprising similar battery configuration (as \emph{agent A} in Case 1) and having similar \emph{reservation values}.
Figure~\ref{fig_nash_distace} elucidates the quality of the outcome --i.e. \emph{agreement} -- through the distribution of the Euclidean distance from the \emph{agreement} to the \emph{Nash solution}, and how the \emph{agreement} outperforms the \emph{no-deal solution} by being more likely to be the \emph{Nash solution}. The distances are measured for each negotiation session
 over the whole simulation periods and normalized within unit range. 
 % Apparently, the density of the distribution is at the peak closer to the distance \rm{0.0}, which means the \emph{Negotiation process} is able to find the \emph{agreements} that are the \emph{Nash solution} most of the time. 
% \todo{Mathematical interpretation? E.g. a social regret measure (loss of social welfare) wrt an optimal offer / Pareto front} As shown in the figure, most of the agreed offers are located closer to the the \emph{Nash solution}. \todo{which one? Nash point (product) or social welfare (sum)? You could evaluate this in different ways and use both.}.
Now, we turn the analysis toward the allocative efficiency of the proposed negotiation strategy, and how the strategy establishes itself preferable for all agents over a baseline strategy of \emph{no flexibility} and a strategy of \emph{individual control} of flexibility without any P2P exchange. Figure~\ref{fig_improvement_sw} presents the relative increase in utility for each prosumer of an EC with 9 agents, comparing the improvements of \emph{individual control} strategy and \emph{negotiation and control} strategy over \emph{No flexibility} strategy. As seen in the figure, $(u_{i}(s_2)-u_{i}(s_0))$ dominates over $(u_{i}(s_1)-u_{i}(s_0))$ by placing itself over the dashed line. Therefore, it implies that the \emph{social welfare} criteria, both \emph{utilitarian} and \emph{Nash} are maximized by the proposed \emph{negotiation and control} strategy. 

% \comments{Take-away message}{(also put a shorter and reworded version in the intro / conclusion) This last result provides argument for out main statement. Current energy systems still have silos. Has to make use of untapped synergies. Peer to peer negotiation can in fact provide alleviation for this setting and increase social welfare, already building from a single pairwise interaction. (Spin it towards market people; wag the dog)}

\section{Conclusion}
\label{sec_conclusion}
A residential cooperative potentially exhibits inefficiencies due to renewable power integration and uncoordinated activations of locally owned distributed energy resources of heterogeneous prosumers.
Automated negotiation -- a natural model of interaction -- has the ability to alleviate these inefficiencies by accommodating the heterogeneous preferences of prosumers in joint decision making. In this paper, we have presented a P2P automated \emph{bilateral negotiation} strategy for \emph{energy contract} settlement between prosumers. The prosumers jointly seek for an \emph{agreement} on \emph{energy contracts/loans} -- consisting of energy volume to be exchanged and the return time of the exchanged energy -- that maximise their preferences by evaluating the realized energy profiles and the consequent flexibility dispatch. Although we consider a predefined set of criteria for the agents to have the preferences on, in reality, the agents may have a diverse set of mutually exclusive constraints that shape up their personal preferences. The proposed negotiation strategy allows the agents to effortlessly stack-up those local constraints weighing by preferences while settling for the \emph{contracts}.
The proposed negotiation based strategy is applied to real energy profiles, and results in an improved \emph{utilitarian} social welfare as well as improved fairness w.r.t. \emph{Nash} social welfare; which is remarkable considering that the allocations are achieved from single pairwise interactions amongst prosumers. 

In this paper, we assume the weights an agent places on the criteria to be predefined, whereas in practice, an agent may be uncertain about the preferences and may need to elicit them from prosumers in a cost-effective way~\cite{BaarslagVisionary2017,Long2018}. Future work may investigate the case where the agents exhibit uncertainty over the preferences and are required to negotiate successfully with partial preferences.

% Residential energy cooperatives have potential to mitigate the increasing fluctuation due to renewable power generation

% In this paper, we did not consider multiple negotiation sessions to trade residual demand. Moreover, we assume the weights an agent places on the criteria are predefined, whereas an agent may be uncertain about the preferences. \todo{All these are future outlooks: make them positive instead of what we don't do.}

\section*{Acknowledgment}
% This research has received funding through the ERA-Net Smart
% Grids Plus project Grid-Friends, with support from the European Union’s Horizon 2020
% research and innovation programme.
The Fraunhofer Institute for Industrial Mathematics ITWM, Kaiserslautern, has kindly provided power time series data for residential load and PV generation, which is underlying the numerical evaluation of the proposed methodology. % This acknowledgement has been provided by Fraunhofer.
This research has received funding through the ERA-Net Smart Grid Plus project Grid-Friends (with support from the European Union’s Horizon 2020
research and innovation programme) and the Veni research programme with project number 639.021.751, which is financed by the Netherlands Organisation for Scientific Research (NWO).
% Combining both funding information.
% \bibliographystyle{IEEEtran}
% \bibliography{my_ref}

\end{document}